\documentclass[aps,prl,twocolumn,showpacs]{revtex4}
\usepackage{amssymb}
\usepackage{amsmath}
\usepackage{graphicx}
\usepackage{pstricks}
\usepackage{epsfig}

\begin{document}


\def\beq{\begin{equation}}
\def\eeq{\end{equation}}
\newcommand{\Tr}{{\rm Tr}} 
\newcommand{\tr}{{\rm tr}} 
\newcommand{\mean}[1]{\langle #1 \rangle}
\def\eps{\epsilon}
\def\gam{\gamma} 
\def\half{\frac{1}{2}}
\def\p{{\bf p}} 
\def\q{{\bf q}}
\def\r{{\bf r}}
\def\t{{\bf t}}
\def\u{{\bf u}}
\def\v{{\bf v}}
\def\x{{\bf x}}
\def\y{{\bf y}} 
\def\z{{\bf z}} 
\def\A{{\bf A}}
\def\B{{\bf B}}
\def\D{{\bf D}} 
\def\E{{\bf E}} 
\def\F{{\bf F}} 
\def\H{{\bf H}}  
\def\J{{\bf J}}
\def\K{{\bf K}} 

\def\L{{\bf L}}
\def\M{{\bf M}}  
\def\O{{\bf O}} 
\def\P{{\bf P}} 
\def\Q{{\bf Q}} 
\def\R{{\bf R}}
\def\S{{\bf S}}
\def\nablabf{\boldsymbol{\nabla}}

\def\w{\omega}
\def\wn{\omega_n}
\def\wnu{\omega_\nu}
\def\wp{\omega_p} 
\def\dmu{{\partial_\mu}}
\def\dl{{\partial_l}}  
\def\dt{\partial_t}
\def\dx{\partial_x}
\def\dy{\partial_y} 
\def\dtau{{\partial_\tau}}  
\def\det{{\rm det}} 

\def\dsum{\displaystyle \sum}
\def\dint{\displaystyle \int} 
\def\intt{\int_{-\infty}^\infty dt} 
\def\inttp{\int_{-\infty}^\infty dt'} 
\def\intk{\int_{\bf k}} 
\def\intkd{\int \frac{d^dk}{(2\pi)^d}}
\def\intq{\int_{\bf q}} 
\def\intr{\int d^dr}  
\def\dintr{\displaystyle \int d^dr} 
\def\intrp{\int d^dr'}
\def\dinttau{\displaystyle \int_0^\beta d\tau}
\def\dinttaup{\displaystyle \int_0^\beta d\tau'}
\def\inttau{\int_0^\beta d\tau}
\def\inttaup{\int_0^\beta d\tau'}
\def\intx{\int d^{d+1}x} 
\def\inttaur{\int_0^\beta d\tau \int d^dr}
\def\intinf{\int_{-\infty}^\infty}
\def\dinttaur{\displaystyle \int_0^\beta d\tau \int d^dr}
\def\dintinf{\displaystyle \int_{-\infty}^\infty}
\def\intw{\int_{-\infty}^\infty \frac{d\w}{2\pi}}
\def\calO{{\cal O}}


\title{Unified picture of superfluidity: From Bogoliubov's approximation to Popov's hydrodynamic theory} 

\author{N. Dupuis}
\affiliation{
 Laboratoire de Physique Th\'eorique de la Mati\`ere Condens\'ee, 
CNRS UMR 7600, \\ Universit\'e Pierre et Marie Curie, 4 Place Jussieu, 
75252 Paris Cedex 05,  France}
\affiliation{Laboratoire de Physique des Solides, CNRS UMR 8502, Universit\'e Paris-Sud, 91405 Orsay, France}

\date{April 21, 2009}
\begin{abstract} 
Using a non-perturbative renormalization-group technique, we compute the momentum and frequency dependence of the anomalous self-energy and the one-particle spectral function of two-dimensional interacting bosons at zero temperature. Below a characteristic momentum scale $k_G$, where the Bogoliubov approximation breaks down, the anomalous self-energy develops a square root singularity and the Goldstone mode of the superfluid phase (Bogoliubov sound mode) coexists with a continuum of excitations, in agreement with the predictions of Popov's hydrodynamic theory. Thus our results provide a unified picture of superfluidity in interacting boson systems and connect Bogoliubov's theory (valid for momenta larger than $k_G$) to Popov's hydrodynamic approach. 
\end{abstract}

\maketitle

Following the success of the Bogoliubov theory~\cite{Bogoliubov47} in providing a microscopic explanation of superfluidity, much theoretical work has been devoted to the calculation of the one-particle Green function of Bose superfluids~\cite{note5}. Early attempts to improve the Bogoliubov approximation however encountered difficulties due to a singular perturbation theory plagued by infrared divergences~\cite{Gavoret64}. In the 70s, Nepomnyashchii and Nepomnyashchii proved that the anomalous self-energy (the main quantity determining the one-particle Green function) vanishes at zero frequency and momentum in dimension $d\leq 3$~\cite{Nepomnyashchii75,Nepomnyashchii78}. This exact result shows that the Bogoliubov approximation, where the linear spectrum and the superfluidity rely on a finite value of the anomalous self-energy, breaks down at low energy. The vanishing of the anomalous self-energy in the infrared limit has a definite physical origin; it is due to the coupling between transverse (phase) and longitudinal fluctuations and the resulting divergence of the longitudinal susceptibility -- a general phenomenon in systems with a continuous broken symmetry~\cite{Patasinskij73}. 

An alternative approach to superfluidity, based on a phase-amplitude representation of the boson field, has been proposed by Popov~\cite{Popov_book}. This approach is free of infrared singularity, but restricted to the (low-momentum) hydrodynamic regime and therefore does not allow to study the higher-momentum regime where the Bogoliubov approximation is valid. Nevertheless, Popov's theory agrees with the exact result and the asymptotic low-energy behavior obtained by Nepomnyashchii and co-workers~\cite{Popov79,Nepomnyashchii83}. Furthermore, the expression of the anomalous self-energy obtained by Nepomnyashchii {\it et al.} and Popov in the low-energy limit yields a continuum of (one-particle) excitations coexisting with the Bogoliubov sound mode~\cite{Giorgini92}, in marked contrast with the Bogoliubov theory where the sound mode is the sole excitation at low energies.  

Although the breakdown of the Bogoliubov approximation in $d\leq 3$ is now well understood within the renormalization group approach~\cite{Pistolesi04,Wetterich08,Dupuis07,Floerchinger08}, no theoretical framework has given a unified description, from high to low energies, of superfluidity in interacting boson systems. Taking advantage of recent progress in the non-perturbative renormalization group (NPRG), we have calculated the momentum and frequency dependence of the anomalous self-energy of interacting bosons. Our results provide a unified description of superfluidity that encompasses both Bogoliubov's theory and Popov's hydrodynamic approach. 

In this Letter, we focus on a two-dimensional system~\cite{note6}. We show that the Bogoliubov approximation breaks down at a characteristic momentum scale $k_G$ which, for weak boson-boson interactions, is much smaller than the inverse healing length $k_h$~\cite{note2}. Moreover, the anomalous self-energy $\Sigma_{\rm an}(\p,\w)$ becomes singular below $k_G$ and induces a continuum of excitations which coexists with the Bogoliubov sound mode, in agreement with the predictions of Popov's hydrodynamic approach. We compute the longitudinal (one-particle) spectral function $A_{\rm ll}(\p,\w)$ and discuss its dependence on $|\p|/k_G$~\cite{note7}. 

We consider two-dimensional interacting bosons at zero temperature, with the action
\beq
S = \int dx \left[ \psi^*(x)\left(\dtau-\mu - \frac{\nablabf^2}{2m}
  \right) \psi(x) + \frac{g}{2} |\psi(x)|^4 \right] 
\label{action} 
\eeq
($\hbar=k_B=1$ throughout the Letter), where $\psi(x)$ is a bosonic (complex) field, $x=(\r,\tau)$, and $\int dx=\inttau \int d^2r$. $\tau\in [0,\beta]$ is an imaginary time, $\beta\to\infty$ the inverse temperature, and $\mu$ denotes the
chemical potential. The interaction is assumed to be local in space and the
model is regularized by a momentum cutoff $\Lambda$. It is convenient to write the boson field as $\psi=\frac{1}{\sqrt{2}} (\psi_1+i\psi_2)$ with  $\psi_1$ and $\psi_2$ real. We assume the dimensionless coupling parameter $2gm$ to be much smaller than unity (weak coupling limit). 

The excitation spectrum can be obtained from the one-particle Green function 
\begin{align}
G_{ij}(p;\phi) ={}& \frac{\phi_i\phi_j}{2n} G_{\rm ll}(p;n) + \left(\delta_{ij}-\frac{\phi_i\phi_j}{2n} \right) G_{\rm tt}(p;n) \nonumber \\ & + \eps_{ij} G_{\rm lt}(p;n) 
\label{Gdef}
\end{align}
or its inverse, the two-point vertex 
\beq
\Gamma^{(2)}_{ij}(p;\phi) = \delta_{i,j} \Gamma_A(p;n) + \phi_i\phi_j \Gamma_B(p;n) +\eps_{ij} \Gamma_C(p;n) ,
\label{Gam2def}
\eeq
where $\eps_{ij}$ is the antisymmetric tensor and $p=(\p,i\w)$ (with $\w$ a Matsubara frequency). $\phi_i=\mean{\psi_i(x)}$ ($i=1,2$) is assumed to be space and time independent, and $n=(\phi_1^2+\phi_2^2)/2$ denotes the condensate density. Taking advantage of the (global) gauge invariance of the action (\ref{action}), we have introduced different Green functions for transverse (phase) and longitudinal fluctuations wrt the constant field $(\phi_1,\phi_2)$. Formally, the two-point vertex $\Gamma^{(2)}_{ij}(p;\phi)$ can be defined as the second-order functional derivative of the effective action $\Gamma[\phi]$ (the generating functional of one-particle irreducible vertices). The NPRG procedure is set up by adding to the action (\ref{action}) an infrared regulator $\Delta S_k=\sum_p \psi^*(p)R_k(\p)\psi(p)$ which suppresses fluctuations with momenta below a characteristic scale $k$ but leaves the high-momenta modes unaffected (for a review, see Ref.~\cite{Berges00}). The dependence of the effective action on $k$ is given by Wetterich's equation~\cite{Wetterich93}
\beq
\partial_k \Gamma_k[\phi] = \half \Tr \left\lbrace \partial_k
R_k\bigl(\Gamma^{(2)}_k[\phi]+R_k\bigr)^{-1}\right\rbrace .
\label{Weq} 
\eeq
In Fourier space, the trace involves a sum over frequencies and momenta as well as a trace over the two components of the field $\phi=(\phi_1,\phi_2)$. For $k\simeq \Lambda$, the regulator $\Delta S_k[\phi]$ suppresses fluctuations and the mean-field theory, where the effective action $\Gamma_\Lambda[\phi]$ reduces to the microscopic action $S[\phi]$, becomes exact thus reproducing the results of the Bogoliubov approximation. On the other hand for $k=0$, provided that $R_k(\p)$ vanishes, $\Gamma_k[\phi]$ gives the effective action of the original model (\ref{action}) from which we can deduce $\Gamma^{(2)}$ and the one-particle Green function. 

We make the following two simplifications: i) we neglect the field dependence of $\Gamma_{\alpha,k}(p;n)$ ($\alpha=A,B,C$) which we approximate by its value at the actual ($k$-dependent) condensate density $n_{0,k}$. $n_{0,k}$ is obtained from the minimum of the effective potential $U_k(n)=(\beta V)^{-1}\Gamma_k[\phi]$ with $\phi=(\sqrt{2n},0)$ ($V$ is the volume of the system). ii) We approximate the effective potential by 
\beq
U_k(n) = U_k(n_{0,k}) + \frac{\lambda_k}{2}(n-n_{0,k})^2 ,
\eeq
where $\lambda|_{k=\Lambda}=g$. Our approach follows the NPRG scheme recently proposed by Blaizot, M\'endez-Galain and Wschebor and others~\cite{Blaizot06,Benitez08,Benitez09} with a truncation in fields to lowest non-trivial order~\cite{Guerra07}. 

\begin{figure}
\centerline{\includegraphics[width=7.5cm,clip]{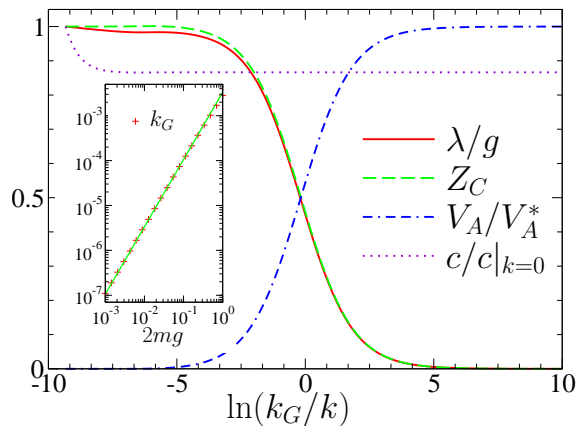}}
\caption{(Color online) $\lambda/g$, $Z_C$, $V_A/V_A^*$ and $c/c|_{k=0}$ {\it vs.} $\ln(k_G/k)$, where $k_G=\sqrt{(gm)^3\bar n}/4\pi$, for $\bar n=0.01$ and $2mg=0.1$ ($\ln(k_G/k_h)\simeq -5.87$). The inset shows $k_G$ {\it vs.} $2mg$ obtained from the criterion $V_A|_{k_G}=V_A^*/2$ (the green solid line is a fit to $k_G\propto (2mg)^{3/2}$). All figures are obtained with $\Lambda=1$, $2m=1$ and the regulator $R_k(\p)=Z_{A,k} \eps_\p/(e^{\p^2/k^2}-1)$. }
\label{fig_de}
\end{figure}

Previous NPRG studies of interacting bosons assumed a simple form of the effective action $\Gamma_k[\phi]$ with local and $\calO(\nablabf^2,\dtau,\partial_\tau^2)$ terms~\cite{Wetterich08,Dupuis07,Floerchinger08}. In our formalism, this amounts to expanding the vertices $\Gamma_A$, $\Gamma_B$, $\Gamma_C$ in powers of $\p$ and $\w$,
\beq
\Gamma_A(p) = V_A \w^2+Z_A \eps_\p , \;\;
\Gamma_B(p) = \lambda , \;\;
\Gamma_C(p) = Z_C\w 
\label{Gamde}
\eeq 
(we drop the $k$ index to alleviate the notation), where $\eps_\p=\p^2/2m$ is the dispersion of the free bosons. The initial conditions ($k=\Lambda$), $Z_A=Z_C=1$, $V_A=0$, $\lambda=g$ and $n_0=\mu/g$, reproduce the Green function $G_{ij}(p)$ of the Bogoliubov approximation. Equations (\ref{Gdef},\ref{Gamde}) yield a low-energy mode $\w=c|\p|$, with ($k$-dependent) velocity 
\beq
c = \left( \frac{Z_A/(2m)}{V_A+Z_C^2/(2\lambda n_0)} \right)^{1/2} ,
\eeq
and a superfluid density $n_s=Z_An_0=\bar n$ where $\bar n$ is the mean boson density~\cite{Dupuis07,Floerchinger08}. In the weak-coupling limit, $n_0$ and $Z_A$ are weakly renormalized and $n_0^*\simeq \mu/g$, $Z_A^*\simeq 1$ (with $n_0^*=n_0|_{k=0}$, etc.). On the other hand $Z_C\sim k$ and $\lambda\sim k$ vanish when $k\to 0$ while $V_A$ takes a finite value $V_A^*$ (Fig.~\ref{fig_de})~\cite{Wetterich08,Dupuis07}. The anomalous self-energy $\Sigma_{\rm an}(0,0)=n_0\Gamma_B(0,0)=n_0 \lambda$ therefore vanishes for $k\to 0$ in agreement with the exact result~\cite{Nepomnyashchii75}. The existence of a linear spectrum is then due to the relativistic form of the action which emerges at low energy ($Z_C\to 0$ and $V_A\to V_A^*>0$)~\cite{Wetterich08,Dupuis07}. The characteristic momentum scale $k_G$ (``Ginzburg'' scale) at which the Bogoliubov approximation breaks down can be defined by the criterion $V_A|_{k_G}=V_A^*/2$. In the weak coupling limit $2gm\ll 1$, it is found to be proportional to $\sqrt{(gm)^3\bar n}\sim gm k_h$ ($k_h=\sqrt{2mg\bar n}$ is the inverse healing length below which the spectrum becomes linear) in agreement with a simple estimate based on the (one-loop) perturbative correction to the Bogoliubov approximation~\cite{Pistolesi04} (see the inset in Fig.~\ref{fig_de}). In practice, we use the definition $k_G=\sqrt{(gm)^3\bar n}/(4\pi)$. 

\begin{figure}[t]
\centerline{\includegraphics[clip,width=7.5cm]{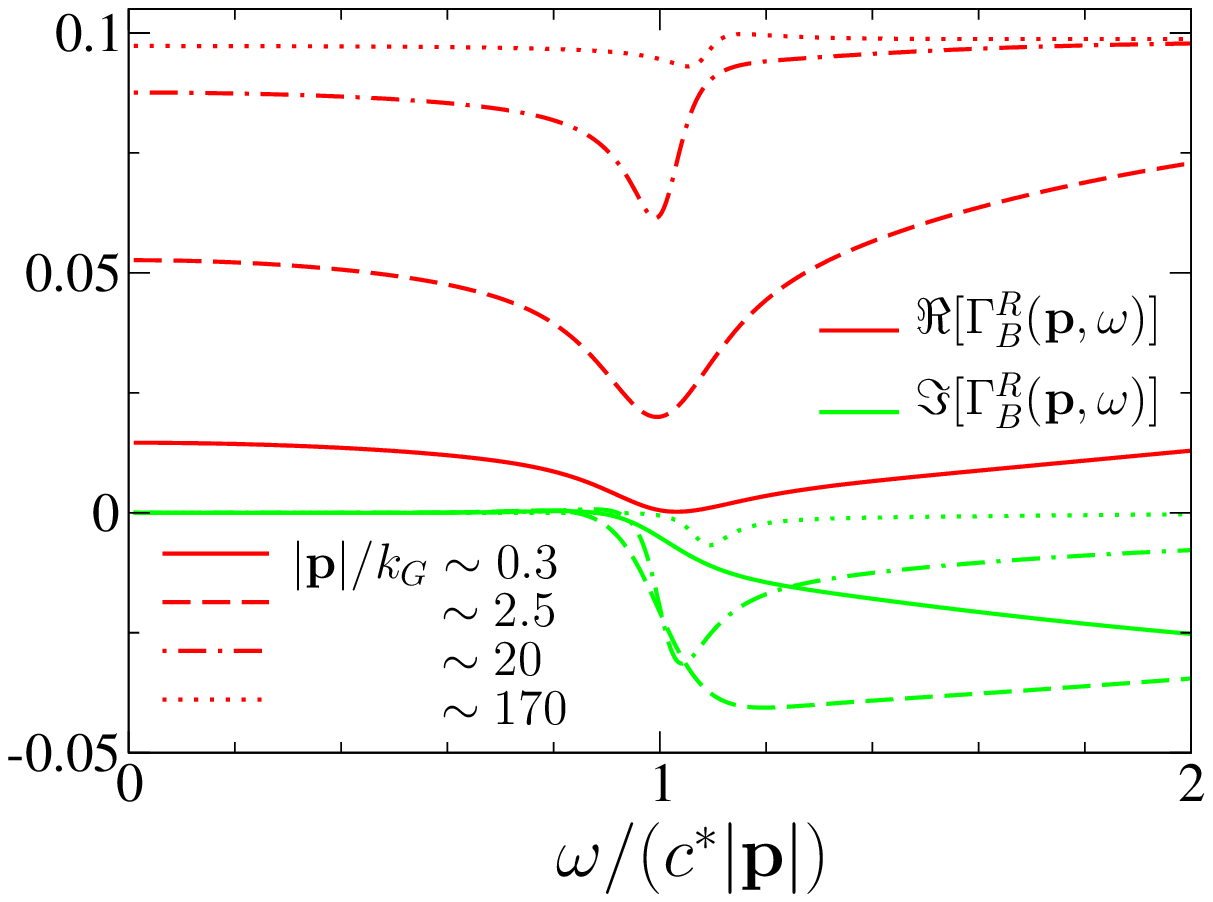}}
\caption{(Color online) Real and imaginary parts of the retarded vertex $\Gamma^R_B(\p,\w)=\Sigma^R_{\rm an}(\p,\w)/n_0$ for various values of $|\p|$ ranging from $0.3 k_G$ up to $170k_G\sim k_h/2$ ($\bar n=0.01$, $2mg=0.1$ and $k=0$). The Bogoliubov approximation corresponds to $\Gamma_B^R(\p,\w)=g=0.1$. ($\Re[\Gamma_B^R]\geq 0$ and  $\Im[\Gamma_B^R]\leq 0$.)}
\label{fig_gamB_1}
\centerline{\includegraphics[clip,width=7.5cm]{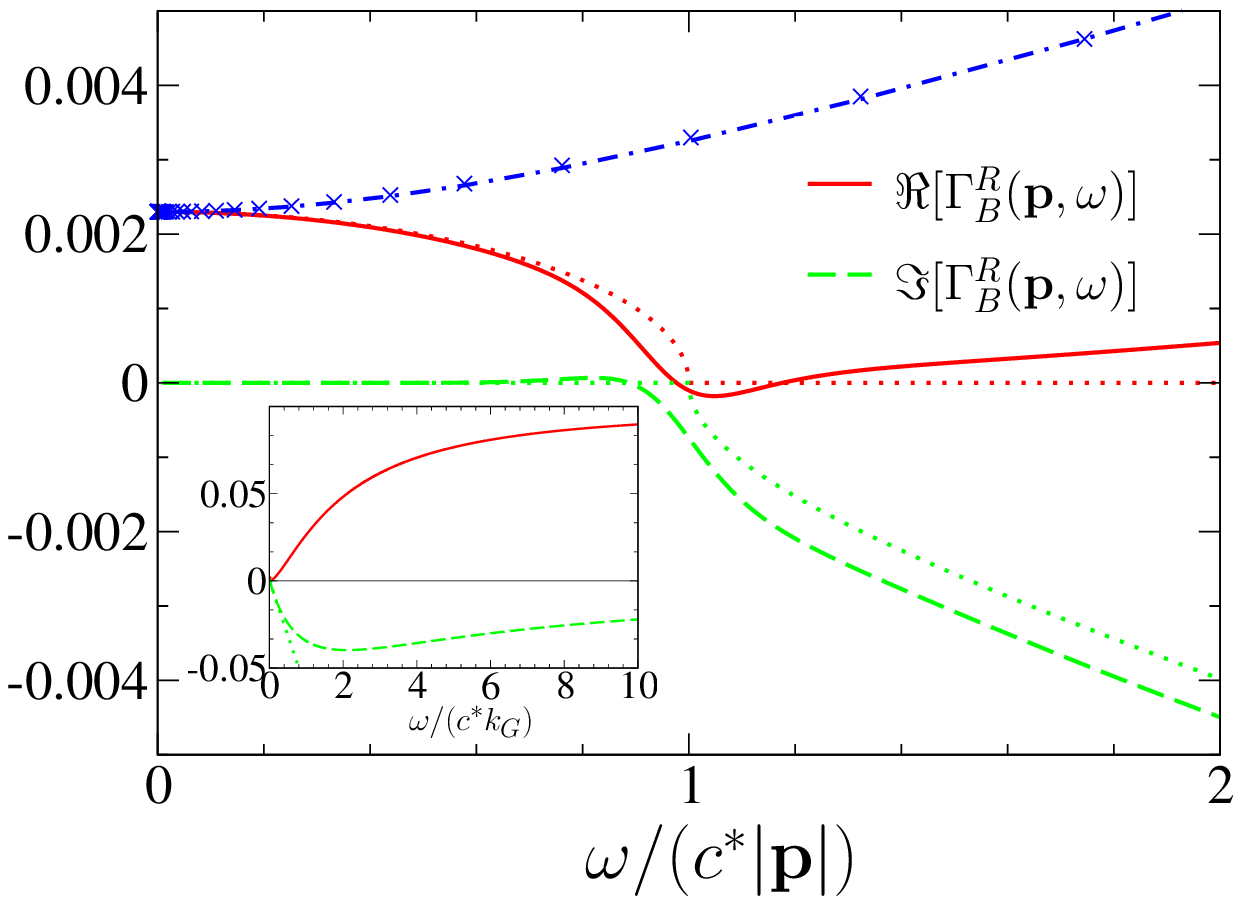}}
\caption{(Color online) Same as figure \ref{fig_gamB_1} but for $|\p|/k_G\simeq 0.036$. The dotted lines show the analytical result (\ref{fit_R}) obtained from the approximation (\ref{fit_B}). The latter is shown by the blue dashed-dotted line while the blue crosses give the numerical results for $\Gamma_B(\p,i\w)$ {\it vs.} the rescaled Matsubara frequency $\w/(c^*|{\bf p}|)$. The inset shows $\Gamma_B^R(\p,\w)$ on a much larger energy scale ($\omega\leq 10c^*k_G\sim 280 c^*|\p|$; the square root singularity at $\w\ll c^*k_G$ is not visible on this plot).}
\label{fig_gamB_2}
\end{figure} 

Now we discuss the momentum and frequency dependence of the vertices beyond their derivative expansion (\ref{Gamde}). To simplify the numerical evaluation of $\partial_k \Gamma_{\alpha,k}(p)$ ($\alpha=A,B,C$), we approximate the propagators entering the flow equations using (\ref{Gamde}). This type of approximation has been shown to be very accurate for classical systems~\cite{note1}. It is sufficient to obtain the anomalous self-energy $\Sigma_{\rm an}(p)=n_0\Gamma_B(p)$ but is less reliable for the calculation of the (small) damping terms arising from $\Gamma_A$ and $\Gamma_C$~\cite{note3}. In practice, we compute the vertices $\Gamma_\alpha(\p,i\w)$ for typically 100 frequency points and then use a Pad\'e approximant to obtain the retarded part $\Gamma^R_\alpha(\p,\w)=\Gamma_\alpha(\p,\w+i0^+)$~\cite{Vidberg77}. While the Bogoliubov result $\Gamma^R_B(\p,\w)=g$ is a good approximation when $|\p|\gg k_G$, $\Gamma_B^R(\p,\w)$ develops a strong frequency dependence for $|\p|\lesssim k_G$ (Fig.~\ref{fig_gamB_1}). In the limit $|\p|\ll k_G$ and $|\w|\ll c^*k_G$, the vertices $\Gamma_A$ and $\Gamma_C$ are very well approximated by their low-energy limit (\ref{Gamde}), while $\Gamma_B(p)$ is well fitted by a square-root singularity,
\beq
\Gamma_B(\p,i\w) \simeq C \sqrt{\w^2+(c^*\p)^2} ,
\label{fit_B} 
\eeq
with $C$ a $\p$-independent constant (the Bogoliubov result $\Gamma^R_B(\p,\w)\simeq g$ is nevertheless reproduced at high energies $|\w|\gg c^*k_G$) (Fig.~\ref{fig_gamB_2}). For the retarded part, equation (\ref{fit_B}) implies 
\begin{align}
\Gamma_B^R(\p,\w) \simeq {}& C \theta(c^*|\p|-\w) \sqrt{(c^*\p)^2-\w^2} 
\nonumber \\ & - i C \theta(\w-c^*|\p|) \sqrt{\w^2-(c^*\p)^2}  
\label{fit_R} 
\end{align}
($\theta(x)$ is the step function) when $|\p|,\w/c^*\ll k_G$ (we discuss only $\w\geq 0$). The square root singularity (\ref{fit_R}) of the anomalous self-energy $n_0\Gamma_B^R(\p,\w)$ agrees with the result obtained from diagrammatic resummations~\cite{Nepomnyashchii78} or the predictions of the hydrodynamic approach~\cite{Popov79} in the limit $(\p,\w)\to 0$. As shown in Fig.~\ref{fig_gamB_2}, this singularity is very well reproduced by the result deduced from the Pad\'e approximant. Thus our results interpolate between Bogoliubov's approximation ($|\p|\gg k_G$) and Popov's hydrodynamic theory ($|\p|\ll k_G$). 

\begin{figure}[t]
\centerline{\includegraphics[clip,width=6.5cm]{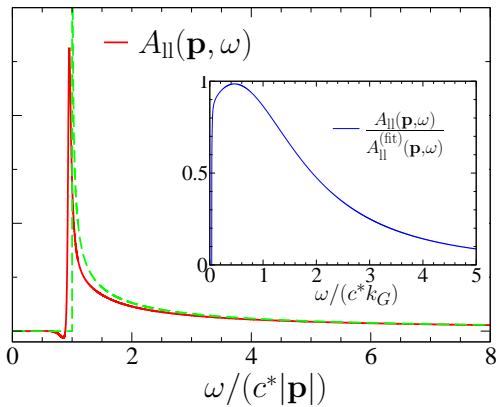}}
\caption{(Color online) Longitudinal spectral function $A_{\rm ll}(\p,\w)$ for $|\p|/k_G\simeq 0.036$ and the same parameters as in figure \ref{fig_gamB_2}. The red solid line is the result obtained from the Pad\'e approximant, while the green dashed line is obtained from the analytic expression (\ref{fit_All}). The inset shows the ratio between $A_{\rm ll}(\p,\w)$ and the approximate result (\ref{fit_All}) on a larger energy scale.}
\label{fig_All}
\end{figure}

In the low-energy limit $\p,\w\to 0$, $\Gamma_B^R(\p,\w)$ is of order $|\p|,|\w|$, while $\Gamma_A^R(\p,\w)$ and $\Gamma_C^R(\p,\w)^2$ are of order $\p^2,\w^2$. The one-particle Green function then becomes
\begin{align}
G_{\rm tt}^R(\p,\w) &= - \frac{\Gamma^R_A(\p,\w)+2n_0^*\Gamma_B^R(\p,\w)}{D^R(\p,\w)} \simeq - \frac{1}{\Gamma^R_A(\p,\w)} , \nonumber \\ 
G_{\rm ll}^R(\p,\w) &= - \frac{\Gamma_A^R(\p,\w)}{D^R(\p,\w)} \simeq - \frac{1}{2n_0^*\Gamma^R_B(\p,\w)} , \label{GR} \\ 
G^R_{\rm lt}(\p,\w) &= \frac{\Gamma^R_C(\p,\w)}{D^R(\p,\w)} , \nonumber 
\end{align}
where $D^R(\p,\w)=\Gamma^R_A(\p,\w)[\Gamma^R_A(\p,\w)+ 2n_0^*\Gamma_B^R(\p,\w)] +\Gamma_C^R(\p,\w)^2$. From (\ref{Gamde}) and (\ref{GR}), one concludes that the transverse spectral function,
\beq
A_{\rm tt}(\p,\w) = - \frac{1}{\pi} \Im[G^R_{\rm tt}(\p,\w)] \simeq \frac{\delta(\w-c^*|\p|) }{2V_A^*c^*|\p|} 
\eeq
(for $\w\geq 0$), exhibits a Dirac-like peak at the Bogoliubov mode frequency $\w=c^*|\p|$, in very good agreement with the result obtained from the Pad\'e approximant~\cite{note3}. From (\ref{fit_R}) and (\ref{GR}), we deduce that the longitudinal spectral function
\beq
A_{\rm ll}(\p,\w) \simeq \frac{1}{2\pi n_0^* C} \frac{\theta(\w-c^*|\p|)}{\sqrt{\w^2-(c^*\p)^2}} 
\label{fit_All}
\eeq
exhibits a continuum of excitations with a singularity at the Bogoliubov mode frequency $\w=c^*|\p|$, again in agreement with the prediction of the hydrodynamic approach~\cite{Giorgini92}. The analytic expression (\ref{fit_All}) gives a good approximation to the result obtained from the Pad\'e approximant when $|\p|\ll k_G$ and $|\w|\ll c^*k_G$ (Fig.~\ref{fig_All}). This defines the domain of validity of the Popov hydrodynamic approach. For $|\p|\sim k_G$, the continuum of excitations is suppressed and the longitudinal spectral function $A_{\rm ll}(\p,\w)$ reduces to a Dirac-like peak as in the Bogoliubov theory.  

It should be noticed that we have retained only the leading term $\Gamma_B(p=0)=\lambda$ in the derivative expansion of $\Gamma_B(p)$ [Eq.~(\ref{Gamde})]. This is made possible by the fact that $\Gamma_B(p=0)=\lambda=\calO(k)$ is a very large energy scale wrt $\Gamma_A(p),\Gamma^2_C(p)\sim k^2$ for typical momentum and frequency $|\p|,\w/c=\calO(k)$. The $(\p,\w)$ dependence of $\Gamma_B(p)\sim \lambda + C\sqrt{\w^2+(c\p)^2}$ does not change the leading behavior of $\Gamma_B(p)=\calO(k)$ which acts as a large mass term in the vertices. It should also be noticed that the singularity of $\Gamma^{(2)}$ yields a similar singularity in higher-order vertices~\cite{Nepomnyashchii75,Nepomnyashchii78}: $\Gamma^{(3)}\sim \sqrt{n_0}\Gamma_B$ and $\Gamma^{(4)}\sim \Gamma_B$. These singular terms are neglected in our approach and we have approximated $\Gamma^{(3)}\sim \sqrt{n_0}\lambda$ and $\Gamma^{(4)}\sim \lambda$. Again this is justified by the fact that $\lambda=\calO(k)$. 

To conclude, we have obtained a unified description of superfluidity which is valid at all energy scales and connect Bogoliubov's theory to Popov's hydrodynamic approach. Our results reveal the fundamental role of the Ginzburg momentum scale $k_G$ in interacting boson systems. $k_G$ sets the scale at which the Bogoliubov approximation breaks down and determines the region $|\p|,|\w|/c^*\ll k_G$ where the longitudinal spectral function $A_{\rm ll}(\p,\w)\sim 1/\sqrt{\w^2-(c^*\p)^2}$ [Eq.~(\ref{fit_All})], i.e. the domain of validity of the hydrodynamic approach. From a more general perspective, our results also show that the NPRG is a very efficient tool to study strongly correlated quantum systems and in particular to compute spectral functions.

\begin{acknowledgments}
The author would like to thank B. Delamotte and N. Wschebor for useful discussions.
\end{acknowledgments}



\begin{thebibliography}{26}
\expandafter\ifx\csname natexlab\endcsname\relax\def\natexlab#1{#1}\fi
\expandafter\ifx\csname bibnamefont\endcsname\relax
  \def\bibnamefont#1{#1}\fi
\expandafter\ifx\csname bibfnamefont\endcsname\relax
  \def\bibfnamefont#1{#1}\fi
\expandafter\ifx\csname citenamefont\endcsname\relax
  \def\citenamefont#1{#1}\fi
\expandafter\ifx\csname url\endcsname\relax
  \def\url#1{\texttt{#1}}\fi
\expandafter\ifx\csname urlprefix\endcsname\relax\def\urlprefix{URL }\fi
\providecommand{\bibinfo}[2]{#2}
\providecommand{\eprint}[2][]{\url{#2}}

\bibitem[{\citenamefont{Bogoliubov}(1947)}]{Bogoliubov47}
\bibinfo{author}{\bibfnamefont{N.~N.} \bibnamefont{Bogoliubov}},
  \bibinfo{journal}{J. Phys. USSR} \textbf{\bibinfo{volume}{11}},
  \bibinfo{pages}{23} (\bibinfo{year}{1947}).

\bibitem[{not({\natexlab{a}})}]{note5}
\bibinfo{note}{For a review, see e.g. H. Shi and A. Griffin, Phys. Rep. {\bf
  304}, 1 (1998).}

\bibitem[{\citenamefont{Gavoret and Nozi\`eres}(1964)}]{Gavoret64}
\bibinfo{author}{\bibfnamefont{J.}~\bibnamefont{Gavoret}} \bibnamefont{and}
  \bibinfo{author}{\bibfnamefont{P.}~\bibnamefont{Nozi\`eres}},
  \bibinfo{journal}{Ann. Phys. (N.Y.)} \textbf{\bibinfo{volume}{28}},
  \bibinfo{pages}{349} (\bibinfo{year}{1964}).

\bibitem[{\citenamefont{Nepomnyashchii and
  Nepomnyashchii}(1975)}]{Nepomnyashchii75}
\bibinfo{author}{\bibfnamefont{A.~A.} \bibnamefont{Nepomnyashchii}}
  \bibnamefont{and} \bibinfo{author}{\bibfnamefont{Y.~A.}
  \bibnamefont{Nepomnyashchii}}, \bibinfo{journal}{JETP Lett.}
  \textbf{\bibinfo{volume}{21}}, \bibinfo{pages}{1} (\bibinfo{year}{1975}).

\bibitem[{\citenamefont{Nepomnyashchii and
  Nepomnyashchii}(1978)}]{Nepomnyashchii78}
\bibinfo{author}{\bibfnamefont{Y.~A.} \bibnamefont{Nepomnyashchii}}
  \bibnamefont{and} \bibinfo{author}{\bibfnamefont{A.~A.}
  \bibnamefont{Nepomnyashchii}}, \bibinfo{journal}{Sov. Phys. JETP}
  \textbf{\bibinfo{volume}{48}}, \bibinfo{pages}{493} (\bibinfo{year}{1978}).

\bibitem[{\citenamefont{Patasinskij and Pokrovskij}(1973)}]{Patasinskij73}
\bibinfo{author}{\bibfnamefont{A.~Z.} \bibnamefont{Patasinskij}}
  \bibnamefont{and} \bibinfo{author}{\bibfnamefont{V.~L.}
  \bibnamefont{Pokrovskij}}, \bibinfo{journal}{Sov. Phys. JETP}
  \textbf{\bibinfo{volume}{37}}, \bibinfo{pages}{733} (\bibinfo{year}{1973}).

\bibitem[{\citenamefont{Popov}(1987)}]{Popov_book}
\bibinfo{author}{\bibfnamefont{V.~N.} \bibnamefont{Popov}},
  \emph{\bibinfo{title}{Functional Integrals and Collective Excitations}}
  (\bibinfo{publisher}{Cambridge University Press},
  \bibinfo{address}{Cambridge, England}, \bibinfo{year}{1987}).

\bibitem[{\citenamefont{Popov and Seredniakov}(1979)}]{Popov79}
\bibinfo{author}{\bibfnamefont{V.~N.} \bibnamefont{Popov}} \bibnamefont{and}
  \bibinfo{author}{\bibfnamefont{A.~V.} \bibnamefont{Seredniakov}},
  \bibinfo{journal}{Sov. Phys. JETP} \textbf{\bibinfo{volume}{50}},
  \bibinfo{pages}{193} (\bibinfo{year}{1979}).

\bibitem{Nepomnyashchii83} Yu.~A. Nepomnyashchii, Sov. Phys. JETP {\bf 58}, 722 (1983); Yu. A. Nepomnyashchii and L.~P. Pitaevskii, Physica B {\bf 194-196}, 543 (1994). 

\bibitem[{\citenamefont{Giorgini et~al.}(1992)\citenamefont{Giorgini,
  Pitaevskii, and Stringari}}]{Giorgini92}
\bibinfo{author}{\bibfnamefont{S.}~\bibnamefont{Giorgini}},
  \bibinfo{author}{\bibfnamefont{L.}~\bibnamefont{Pitaevskii}},
  \bibnamefont{and}
  \bibinfo{author}{\bibfnamefont{S.}~\bibnamefont{Stringari}},
  \bibinfo{journal}{Phys. Rev. B} \textbf{\bibinfo{volume}{46}},
  \bibinfo{pages}{6374} (\bibinfo{year}{1992}).

\bibitem{Pistolesi04} F. Pistolesi {\it et al.}, Phys. Rev. B {\bf 69}, 024513 (2004). 

\bibitem[{\citenamefont{Wetterich}(2008)}]{Wetterich08}
\bibinfo{author}{\bibfnamefont{C.}~\bibnamefont{Wetterich}},
  \bibinfo{journal}{Phys. Rev. B} \textbf{\bibinfo{volume}{77}},
  \bibinfo{eid}{064504} (\bibinfo{year}{2008}).

\bibitem[{\citenamefont{Dupuis and Sengupta}(2007)}]{Dupuis07}
\bibinfo{author}{\bibfnamefont{N.}~\bibnamefont{Dupuis}} \bibnamefont{and}
  \bibinfo{author}{\bibfnamefont{K.}~\bibnamefont{Sengupta}},
  \bibinfo{journal}{Europhys. Lett.} \textbf{\bibinfo{volume}{80}},
  \bibinfo{pages}{50007} (\bibinfo{year}{2007}).

\bibitem[{\citenamefont{Floerchinger and Wetterich}(2008)}]{Floerchinger08}
\bibinfo{author}{\bibfnamefont{S.}~\bibnamefont{Floerchinger}}
  \bibnamefont{and}
  \bibinfo{author}{\bibfnamefont{C.}~\bibnamefont{Wetterich}},
  \bibinfo{journal}{Phys. Rev. A} \textbf{\bibinfo{volume}{77}},
  \bibinfo{eid}{053603} (\bibinfo{year}{2008}).

\bibitem[{not({\natexlab{b}})}]{note6}
\bibinfo{note}{Our results are easily extended to three-dimensional systems. In
  the weak-coupling limit (considered in this Letter), the Ginzburg scale $k_G$
  is however exponentially small wrt the healing scale $k_h$ ($k_G$ and $k_h$
  are defined in the text).}

\bibitem[{not({\natexlab{c}})}]{note2}
\bibinfo{note}{The characteristic momentum scale $k_G$ was first obtained by
Popov and Seredniakov in three dimension~\cite{Popov79}, but its true physical meaning was only discovered later on~\cite{Pistolesi04}.}

\bibitem[{not({\natexlab{d}})}]{note7}
\bibinfo{note}{A calculation of the spectral function of two-dimensional
interacting bosons has recently been reported, with a focus on the
damping of the sound mode, but the longitudinal spectral function has
not been discussed: A. Sinner, N. Hasselmann, and P. Kopietz, Phys. Rev. Lett. {\bf 102}, 120601 (2009).}

\bibitem[{\citenamefont{Berges et~al.}(2000)\citenamefont{Berges, Tetradis, and
  Wetterich}}]{Berges00}
\bibinfo{author}{\bibfnamefont{J.}~\bibnamefont{Berges}},
  \bibinfo{author}{\bibfnamefont{N.}~\bibnamefont{Tetradis}}, \bibnamefont{and}
  \bibinfo{author}{\bibfnamefont{C.}~\bibnamefont{Wetterich}},
  \bibinfo{journal}{Phys. Rep.} \textbf{\bibinfo{volume}{363}},
  \bibinfo{pages}{223} (\bibinfo{year}{2000}).

\bibitem[{\citenamefont{Wetterich}(1993)}]{Wetterich93}
\bibinfo{author}{\bibfnamefont{C.}~\bibnamefont{Wetterich}},
  \bibinfo{journal}{Phys. Lett. B} \textbf{\bibinfo{volume}{301}},
  \bibinfo{pages}{90} (\bibinfo{year}{1993}).

\bibitem[{\citenamefont{Blaizot et~al.}(2006)\citenamefont{Blaizot,
  M\'endez-Galain, and Wschebor}}]{Blaizot06}
\bibinfo{author}{\bibfnamefont{J.-P.} \bibnamefont{Blaizot}},
  \bibinfo{author}{\bibfnamefont{R.}~\bibnamefont{M\'endez-Galain}},
  \bibnamefont{and} \bibinfo{author}{\bibfnamefont{N.}~\bibnamefont{Wschebor}},
  \bibinfo{journal}{Phys. Lett. B} \textbf{\bibinfo{volume}{632}},
  \bibinfo{pages}{571} (\bibinfo{year}{2006}).

\bibitem[{\citenamefont{Benitez et~al.}(2008)\citenamefont{Benitez,
  M\'{e}ndez-Galain, and Wschebor}}]{Benitez08}
\bibinfo{author}{\bibfnamefont{F.}~\bibnamefont{Benitez}},
  \bibinfo{author}{\bibfnamefont{R.}~\bibnamefont{M\'{e}ndez-Galain}},
  \bibnamefont{and} \bibinfo{author}{\bibfnamefont{N.}~\bibnamefont{Wschebor}},
  \bibinfo{journal}{Phys. Rev. B} \textbf{\bibinfo{volume}{77}},
  \bibinfo{pages}{024431} (\bibinfo{year}{2008}).

\bibitem{Benitez09} F. Benitez {\it et al.}, arXiv:0901.0128.

\bibitem[{\citenamefont{Guerra et~al.}(2007)\citenamefont{Guerra,
  M\'endez-Galain, and Wschebor}}]{Guerra07}
\bibinfo{author}{\bibfnamefont{D.}~\bibnamefont{Guerra}},
  \bibinfo{author}{\bibfnamefont{R.}~\bibnamefont{M\'endez-Galain}},
  \bibnamefont{and} \bibinfo{author}{\bibfnamefont{N.}~\bibnamefont{Wschebor}},
  \bibinfo{journal}{Eur. Phys. J. B} \textbf{\bibinfo{volume}{59}},
  \bibinfo{pages}{357} (\bibinfo{year}{2007}).

\bibitem[{not({\natexlab{e}})}]{note1}
\bibinfo{note}{See, for instance, N. Dupuis and K. Sengupta, Eur. Phys. J. B
  {\bf 66}, 271 (2008).}

\bibitem[{not({\natexlab{f}})}]{note3}
\bibinfo{note}{The damping terms $\Im[\Gamma^R_A(\p,\w)]$ and
  $\Re[\Gamma^R_C(\p,\w)]$ do not affect the longitudinal spectral function
  $A_{\rm ll}(\p,\w)$ and can therefore safely be ignored. 
}

\bibitem[{\citenamefont{Vidberg and Serene}(1977)}]{Vidberg77}
\bibinfo{author}{\bibfnamefont{H.~J.} \bibnamefont{Vidberg}} \bibnamefont{and}
  \bibinfo{author}{\bibfnamefont{J.~W.} \bibnamefont{Serene}},
  \bibinfo{journal}{J. Low Temp. Phys} \textbf{\bibinfo{volume}{29}},
  \bibinfo{pages}{179} (\bibinfo{year}{1977}).

\end{thebibliography}

\end{document}